\documentclass[11pt]{article}
\usepackage{epsfig,sint,macros,cite}
\begin{document}
\begin{titlepage}
\begin{flushright}
  CERN-TH/2001/157\\
  CPT-2001/P.4211\\
  DESY 01-083\\
  IFIC/01-34\\
  FTUV-010617\\
  LTH\,508
\end{flushright}

\vskip 0.5 cm
\begin{center}
  {\Large\bf 
  Non-perturbative renormalization of the quark 
condensate in Ginsparg-Wilson regularizations  \\[0.5ex] }
\end{center}
\vskip 0.5 cm
\begin{center}
{\large 
     Pilar Hern\'andez$^{\scriptscriptstyle a,}$\footnote{On leave of
     absence from Departamento de F\'{\i}sica Te\'orica, Universidad
     de Valencia, Spain},
     Karl Jansen$^{\scriptscriptstyle a,b}$,
     Laurent Lellouch$^{\scriptscriptstyle c}$ and
     Hartmut Wittig$^{\scriptscriptstyle a,d,}$\footnote{PPARC
     Advanced Fellow}
\vskip 0.5cm
$^{\scriptstyle a}$
CERN, Theory Division,\\
CH-1211 Geneva 23, Switzerland
\vskip 1.5ex
$^{\scriptstyle b}$
NIC/DESY Zeuthen \\
Platanenallee 6, D-15738 Zeuthen, Germany
\vskip 1.5ex
$^{\scriptstyle c}$
Centre de Physique Th\'eorique, Case 907, CNRS Luminy\\
F-13288 Marseille Cedex 9, France
\vskip 1.5ex
$^{\scriptstyle d}$
Division of Theoretical Physics\\
Department of Mathematical Sciences\\
University of Liverpool, Liverpool L69~3BX, UK
\vskip 1.0cm
{\bf Abstract}}
\vskip 0.35ex
\end{center}

We present a method to compute non-perturbatively the renormalization
constant of the scalar density for Ginsparg-Wilson fermions. It
relies on chiral symmetry and is based on a matching of
renormalization group invariant masses at fixed pseudoscalar meson
mass, making use of results previously obtained by the ALPHA
Collaboration for O($a$)-improved Wilson fermions. Our approach is
quite general and enables the renormalization of scalar and
pseudoscalar densities in lattice regularizations that preserve chiral
symmetry and of fermion masses in any regularization. As an
application we compute the non-perturbative factor which relates the
renormalization group invariant quark condensate to its bare
counterpart, obtained with overlap fermions at $\beta=5.85$ in the
quenched approximation.

\vfill

\begin{center}
June 2001
\end{center}

\eject

\vfill

\eject

\end{titlepage}

\hyphenation{fer-mionic}

\section{Introduction \label{sec_intro}}

The understanding of the low-energy sector of QCD is one of the main
goals of lattice simulations of the theory. In particular, the
determination of the scalar quark-antiquark condensate associated with
the spontaneous breaking of chiral symmetry has been the focus of many
recent
studies~\cite{quark:gupta,cond:APE,cond:SCRI1,cond:SCRI2,cond:paperI,
cond:DeGrand,dwf:RBC00}. There are now good prospects for a reliable
and precise calculation of this quantity, owing to recent progress in
two key areas. The first is the formulation of chiral symmetry on the
lattice: it has recently become clear how the familiar consequences of
chiral symmetry in the continuum can be preserved at non-zero lattice
spacing, by employing lattice Dirac operators that satisfy the
Ginsparg-Wilson relation~\cite{chiral:GWR} (for reviews of the subject
see~\cite{Nied_lat98,Martin_lat99,Neu_lat99}). In the context of
spontaneous symmetry breaking this makes it possible to extract the
quark condensate in a conceptually clean manner through a suitable
finite-size scaling analysis~\cite{fss:karl91,cond:paperI}. Lattice
results for the bare, subtracted condensate in infinite volume,
$-\Ssub$, have already been obtained in this way in the quenched
approximation~\cite{cond:paperI,cond:DeGrand}, using Neuberger's (or
overlap) operator~\cite{chiral:ovlp} as a realization of the
Ginsparg-Wilson relation. In order to present an estimate for the
condensate which can be used in phenomenological applications, the
bare lattice result must, of course, be renormalized.

The renormalization of matrix elements of composite operators defined
on the lattice is the second key area where significant progress has
been achieved (for a recent review see \cite{sint_lat00}). A
theoretical framework to address the problem of (in general
scale-dependent) renormalization of lattice operators in a completely
non-perturbative manner has been developed. The main idea is the
introduction of an intermediate renormalization scheme, such as the
Regularization Independent (RI)~\cite{renorm_mom:paper1} and the
Schr\"odinger functional (SF)~\cite{impr:lett} schemes. These
formalisms have already been used successfully to address the
renormalization of quark masses for
Wilson~\cite{quark:GGRT,quark:marti,mbar:pap3}, staggered
\cite{quark:jlqcd_stag} and Domain Wall \cite{dwf:RBC_npren01}
fermions.

The main subject of this paper is the description of a method to
compute non-perturbatively the multiplicative renormalization constant
of the scalar density in fermionic regularizations based on the
Ginsparg-Wilson relation. Our strategy relies on the matching of
renormalization group invariant quark masses at fixed pseudoscalar
meson mass, and avoids the direct formulation of intermediate
renormalization schemes, such as the SF, for Ginsparg-Wilson fermions.
Instead, we make extensive use of the non-perturbative renormalization
factor for quark masses, computed by the ALPHA Collaboration for
O($a$)-improved Wilson fermions \cite{mbar:pap3}. The method is
applicable to the renormalization of scalar and pseudoscalar densities
for discretizations of the Dirac operator that preserve chiral
symmetry at non-zero lattice spacing, and to the renormalization of
quark masses in any fermion discretization. As an example, we
determine non-perturbatively the renormalization factor which links
the bare scalar condensate obtained using the overlap operator to the
renormalization group invariant condensate.

The important issue of quenching will not be adressed here. While the
methods described below can be applied without change to the full
theory, the calculations discussed are sufficiently costly with
present computer and algorithm technology that unquenched calculations
are not yet considered. Our main purpose is to demonstrate the
feasibility of our strategy, and we restrict ourselves to a single
value of the lattice spacing at which the bare subtracted condensate
has been computed previously, corresponding to a bare coupling of
$\beta=5.85$.

As we will describe below in some detail, our approach to the
non-perturbative determination of the renormalization factor for the
scalar density requires the calculation of two-point functions in the
pseudoscalar channel using overlap fermions. These correlation
functions are themselves a rich source of information on low-energy
QCD, and allow us to compute the condensate independently from the
finite-size scaling analysis of~\cite{cond:paperI}. Furthermore, they
serve to calculate the light quark masses as well as the pseudoscalar
decay constant in the chiral limit. The detailed discussion of these
results is deferred to a companion paper~\cite{cond:paperIV}.

The outline of the remainder of this paper is as follows. In
\Sect{sec_strat} we present our strategy for renormalizing the
condensate. A discussion of cutoff effects associated with the use of
an intermediate O($a$)-improved Wilson regularization and the
intrinsic precision of our method is presented
in~\Sect{sec_cutoff}. \Sect{sec_masses} contains a description of our
simulation results, using the overlap operator as well as
O($a$)-improved Wilson fermions at $\beta=5.85$. In \Sect{sec_results}
we present our results for the renormalization factors and compare
them to one-loop perturbation theory. Finally, \Sect{sec_concl}
contains a summary, including an estimate for the renormalized
condensate at $\beta=5.85$. Some details concerning improvement
coefficients for O($a$)-improved Wilson fermions are described in
Appendix~\ref{app1}.

\hyphenation{re-nor-ma-li-za-tion}

\section{Strategy \label{sec_strat}}

In this section we describe the renormalization of the scalar
condensate for overlap fermions. Our strategy relies on the fact that
the renormalization constants for the scalar and pseudoscalar
densities are identical and are equal to the inverse of the
renormalization factor for fermion masses, as can be shown using the
chiral Ward Identities for Ginsparg-Wilson fermions (see,
e.g. ref.~\cite{chiral:AlFoPaVi})
\be
   \zp=\zs=\frac{1}{\zm}.
\label{eq_zszpzm}
\ee
The non-perturbative renormalization of quark masses has been studied
extensively on the lattice. In particular, the relation between the
renormalization group invariant (RGI) mass and its counterpart in the
intermediate SF scheme is known with an accuracy of better than 2\% in
the continuum limit~\cite{mbar:pap1}. Furthermore, the
non-perturbative matching between the SF scheme and O($a$)-improved
Wilson fermions has been performed for a range of bare
couplings~\cite{mbar:pap1}. One possible strategy is then to repeat
this second step for overlap fermions, by evaluating the normalization
condition of ref.~\cite{mbar:pap1}. However, the direct implementation
of the SF scheme for the overlap operator is not straightforward, due
to the inhomogeneous boundary conditions in the time direction, which
are incompatible with the Ginsparg-Wilson relation. We have thus
devised an alternative strategy, which allows us to exploit the
previously obtained non-perturbative relations between quantities
defined in the O($a$)-improved Wilson theory and their RGI
counterparts.

We begin by considering the renormalization constant
$\zM(g_0)$ which relates a bare quark mass, $m(g_0)$, to the
RGI mass $M$ through
\be
   M = \zM(g_0)\,m(g_0),
   \label{eq_zMdef}
\ee
where we have explicitly indicated the dependence on the bare coupling
$g_0$. Note that we have {\it not} specified the fermionic
discretization at this point. In the case of O($a$)-improved Wilson
fermions we have
\be
   M = \zM^{\rm w}(g_0)\,\mw(g_0),
   \label{eq_zMdef_wil}
\ee
where $\mw$ is the current quark mass, and the superscript ``w''
reminds us that this defines $\zM$ for this particular
regularization. One can now write the ratio $M/m(g_0)$ as
\bea
  \frac{M}{m(g_0)} &=& \frac{M}{\mw(g_0^\prime)}\cdot
                       \frac{\mw(g_0^\prime)}{m(g_0)}   \\
                   &=& \zM^{\rm w}(g_0^\prime)\cdot
            \frac{(r_0\,\mw)(g_0^\prime)}{(r_0\,m)(g_0)}.
\eea
Here, $g_0^\prime$ is a value of the bare coupling which may differ
from $g_0$, and in the last line we have also introduced the hadronic
radius $r_0$~\cite{pot:r0} to set the scale. The factor $\zM^{\rm w}$
has already been computed in the quenched approximation for a large
range of couplings \cite{mbar:pap1}. It is then clear that the
relation between the RGI mass $M$ and the bare mass $m(g_0)$ in any
regularization can simply be obtained by determining the values of
$(r_0\,m)$ and $(r_0\,\mw)$ which reproduce a reference value
$x_{\rm{ref}}$ of a chosen observable. A convenient choice is the
pseudoscalar meson mass in units of $r_0$, such that
$(r_0\mps)^2=x_{\rm ref}$. Thus,
\be
\frac{M}{m(g_0)} = \left.\left\{\left[
     \zM^{\rm w}(g_0')\times(r_0\,\mw)(g_0')\right]
     \cdot\frac{1}{(r_0\,m)(g_0)}\right\}
     \right|_{(r_0\,\mps)^2=x_{\rm ref}.}
\label{eq_zMexpr}
\ee
It is now important to realize that the combination
$\zM^{\rm{w}}(g_0^\prime)\times(r_0\,\mw)(g_0^\prime)$ is a
renormalized, dimensionless quantity. We can therefore define the
universal factor $U_{\rm M}$ in the continuum limit as
\be
   U_{\rm M} = \lim_{g_0^\prime\to0}\left\{\zM^{\rm w}(g_0^\prime)
   \times(r_0\,\mw)(g_0^\prime)\right\}
   \Big|_{(r_0\,\mps)^2=x_{\rm ref}.}
\label{eq_umdef}
\ee
This completes our definition of the renormalization factor $\zM$ for
any given fermionic discretization. By combining
eqs.~(\ref{eq_zMdef}), (\ref{eq_zMexpr}) and~(\ref{eq_umdef}) we
obtain
\be
   \zM(g_0) = U_{\rm M}\cdot\frac{1}{(r_0\,m)
                 }\Big|_{(r_0\,\mps)^2=x_{\rm ref}.}
\label{eq_zMfinal}
\ee
At this point all reference to the bare coupling $g_0^\prime$ and the
use of O($a$)-improved Wilson fermions has disappeared, and the only
part that retains an explicit dependence on the lattice regularization
is the bare quark mass in units of $r_0$, $(r_0\,m)$, at the reference
point $x_{\rm ref}$. The only discretization errors that remain are
those associated with the regularization for which $\zM(g_0)$ is
considered. Estimates for $U_{\rm M}$ in the continuum limit are
easily obtained from published results employing O($a$)-improved
Wilson fermions, which greatly facilitates the evaluation of $\zM$ in
any given regularization. We will return to this point
in~\Sect{sec_cutoff}.

Let us now consider a fermionic regularization which preserves chiral
symmetry. In this case the renormalization factor $\zM$
of~\eq{eq_zMfinal} serves not only to renormalize the quark mass, but
also the scalar condensate. For concreteness we choose overlap
fermions and assume that~\eq{eq_zMfinal} has been evaluated for
$m=\mov$, where $\mov$ denotes the bare mass in the massive overlap
Dirac operator given below in \eq{dn}.\,\footnote{Note that for
Ginsparg-Wilson fermions the bare mass which appears in the Lagrangian
is identical to the quark mass defined through the PCAC relation with
the conserved axial current.}
The RGI condensate $\SRGI$ is then obtained from the bare subtracted
condensate computed using the overlap operator, $\Ssub(g_0)$, through
\be
   \SRGI=\frac{1}{\zM(g_0)}\,\Ssub(g_0).
\label{eq_srgidef}
\ee
The RGI condensate $\SRGI$ is a fully non-perturbative
quantity. However, it is traditional to quote the value of the
condensate in a perturbative scheme such as $\msbar$, at some
reference scale $\mu$. The matching of the RGI condensate to that
defined in the $\msbar$ scheme must necessarily be perturbative. The
$\msbar$ condensate $\Sigma_\msbar(\mu)$ is thus given via
\be
   \Sigma_\msbar(\mu)=\frac{1}{\zm(g_0,\mu)}\,\Ssub(g_0),
\label{eq_smsbardef}
\ee
where
\be
   \zm(g_0,\mu) = \frac{\mbar_\msbar(\mu)}{M}\,\zM(g_0).
\label{eq_zmmudef}
\ee
The numerical value for factor $\mbar_\msbar(\mu)/M$ was
obtained in~\cite{mbar:pap3} through numerical integration of the
perturbative renormalization group (RG) functions in the $\msbar$
scheme. For instance, at the commonly used reference scale
$\mu=2\,\GeV$ the integration of the 4-loop RG functions yields
\be
   \frac{\mbar_\msbar(\mu)}{M}=0.72076,\quad\mu=2\,\GeV\ ,
\label{eq_mbaroverM}
\ee
where the error due to the uncertainty in the quenched value of
$\Lambda_\msbar$ is 1.5\%. The conversion to other reference scales is
easily performed using the tabulated values of
${\mbar_\msbar(\mu)}/{M}$ in Table~3 of ref.~\cite{mbar:pap3}.

The relations between the bare and renormalized condensates of
eqs.~(\ref{eq_srgidef}) and~(\ref{eq_smsbardef}) are more conveniently
expressed in terms of renormalization factors $\zshat$ and $\zs$,
which are related to $\zM$ and $\zm$ by
\be
   \zshat(g_0)  = \frac{1}{\zM(g_0)},\qquad
   \zs(g_0,\mu) = \frac{1}{\zm(g_0,\mu)}.
\label{eq_zs_zm}
\ee
Our goal is to renormalize the bare subtracted condensate calculated
in \cite{cond:paperI}, by computing $\zshat$ at
$\beta=6/g_0^2=5.85$. Thus, in the remainder of this paper we will
focus on the determination of $(r_0\,\mov)$ for $\beta=5.85$ at a
suitably chosen reference point $x_{\rm ref}$. However, before
describing the details of this calculation, we discuss some of the
issues surrounding the determination of the universal factor
$U_{\rm{M}}$.

\section{The universal factor $U_{\rm M}$ , cutoff effects
and overall precision \label{sec_cutoff}}

The factor $U_{\rm M}$ of eq.~(\ref{eq_umdef}) is in fact the RGI
quark mass (in units of $r_0$) in the continuum limit for a
degenerate, pseudoscalar meson with $(r_0\mps)^2=x_{\rm ref}$. In
ref.~\cite{mbar:pap3} RGI quark masses were computed for
$(r_0\mps)^2=1.5736$ and~3.0. The first value corresponds to a
degenerate pseudoscalar meson, which has the same mass as the
kaon.\footnote{Using $\mk=495\,\MeV$ and $r_0=0.5\,\fm$ gives
$(r_0\mk)^2=1.5736$.}
This implies that $U_{\rm{M}}=(M_{\rm{s}}+\widehat{M})r_0/2$ for
$x_{\rm ref}=1.5736$, where $M_{\rm s}$ is the RGI strange quark mass
and $\widehat{M}=\frac{1}{2}(M_{\rm u}+M_{\rm d})$. On the other hand,
choosing $(r_0\mps)^2=x_{\rm{ref}}=3.0$ corresponds to a degenerate
meson with a quark mass roughly equal to that of the strange quark,
such that $U_{\rm{M}}\approx M_{\rm{s}}r_0$. In order to explore the
systematics of our procedure more thoroughly we have considered a
third reference point, $x_{\rm ref}=5.0$, and the reasons for this
choice are explained in Section~\ref{sec_masses}.

The results of ref.~\cite{mbar:pap3} are easily converted into
estimates for $U_{\rm M}$ at $x_{\rm ref}=1.5736$, 3.0 and~5.0: an
estimate of $(M_{\rm{s}}+\widehat{M})r_0$ in the continuum limit is
given in eq.~(5.2) of that paper. Furthermore, by evaluating eq.~(5.1)
for $(r_0\mps)^2=3.0$ in conjunction with the results in the last line
of Table~2, one can infer the value of $U_{\rm M}$ at
$x_{\rm{ref}}=3.0$. Finally, using the results in Table~1, the
procedure of~\cite{mbar:pap3} can be repeated in order to determine
$U_M$ for $x_{\rm{ref}}=5.0$. Thus we obtain
\be
   U_{\rm M} = \left\{ \begin{array}{ll}
                       0.181(6), &\quad x_{\rm ref}=1.5736, \\
                       0.349(9), &\quad x_{\rm ref}=3.0, \\
                       0.580(12),&\quad x_{\rm ref}=5.0.
                       \end{array}\right.
\label{eq_UM_ref}
\ee
Therefore, what is left to do in order to obtain a fully
non-perturbative mass renormalization constant $\zM$ in any given
regularization is to calculate $(r_0\,m)$ in that scheme for one (or
all) of these reference values.

\begin{figure}[t]
\centering
\vspace{-2.7cm}
\epsfig{file=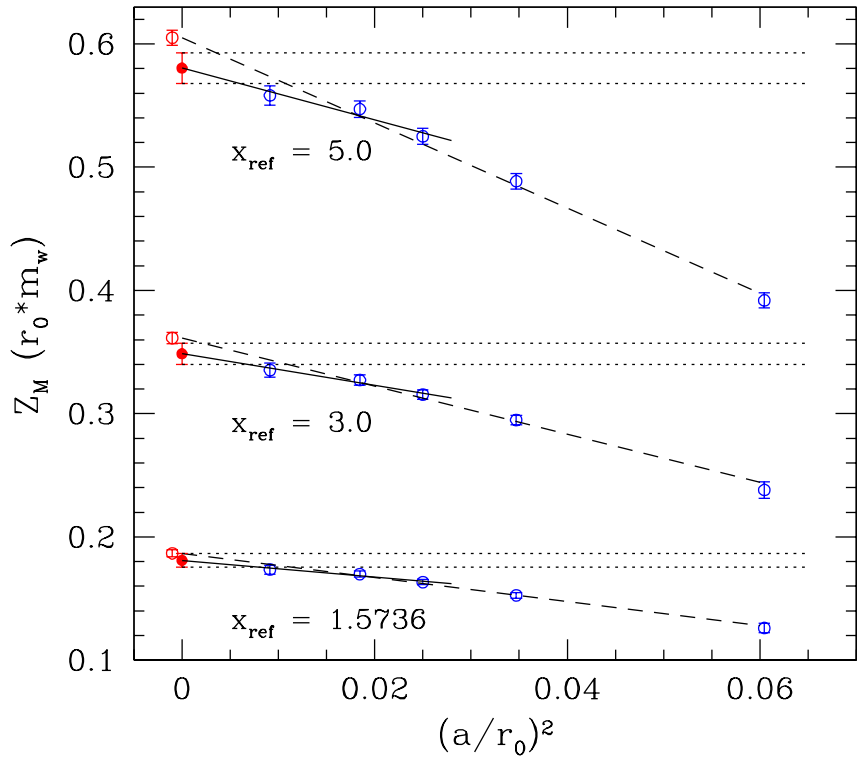, width=15cm} 
\vspace{-2.0cm}
\caption{\it Cutoff effects in $\zM^{\rm w}\,(r_0\,\mw)$ for
$x_{\rm{ref}}=1.5736$, $3.0$ and~$5.0$. The continuum estimates for
$U_{\rm M}$ of \protect\eq{eq_UM_ref} are represented by the error
band (dotted lines). The corresponding linear fits in $(a/r_0)^2$ to
data computed for $\beta\geq6.1$ are shown as the solid
lines. Extrapolations using data at all five values of the lattice
spacings are represented by the dashed lines.}
\label{fig_ZMmw}
\end{figure}

Before undertaking this calculation, we wish to discuss the issue of
discretization errors in our renormalization condition,
\eq{eq_zMfinal}. To this end we note that a valid definition of $\zM$
is also provided by the right-hand side of \eq{eq_zMexpr} evaluated at
non-zero $g_0^\prime$. The two expressions differ by O($a^2$)
discretization errors associated with the intermediate Wilson
regularization. In \eq{eq_zMexpr} these errors are present, while they
are not in \eq{eq_zMfinal}, owing to the continuum extrapolation in
the definition of $U_{\rm M}$. One might think that such O($a^2$)
artefacts -- which are formally of the same or even higher order than
those of most other lattice regularizations -- are unimportant. We are
now going to show that this is not the case. To this end we have
evaluated
\be
   \zM^{\rm w}(g_0^\prime)\times(r_0\mw)(g_0^\prime)
   \Big|_{(r_0\,\mps)^2=x_{\rm ref},} \quad x_{\rm ref}=1.5736,\,3.0,\,5.0
\ee
for bare couplings $g_0^\prime$ corresponding to $\beta=5.85, 6.0,
6.1, 6.2$ and 6.45. Data for $\beta\ge6.0$ were taken from
ref.~\cite{mbar:pap3}. The results for $(r_0\mw)$ at $\beta=5.85$ were
obtained as described in \Sect{sec_masses} below. The value for
$\zM^{\rm w}$ at $\beta=5.85$ was estimated by extrapolating the
parameterization of eq.~(6.10) in~\cite{mbar:pap1} and doubling the
error.

The approach to the continuum limit is shown in
Fig.~\ref{fig_ZMmw}. The first observation is that lattice artefacts
in $\zM^{\rm w}\,(r_0\mw)$ appear to be consistent with the expected
$a^2$ behaviour -- at least for $x_{\rm ref}=1.5736$ and~3.0:
performing linear fits in $(a/r_0)^2$ to data points at all five
values of the lattice spacing produce good
$\chi^2/{\rm{dof}}$. Furthermore, as can be seen from the figure, the
results in the continuum limit are compatible with the estimates for
$U_{\rm M}$ in~\eq{eq_UM_ref}. The latter were obtained by excluding
all points below $\beta=6.1$ from the continuum extrapolations, as a
safeguard against potentially large cutoff effects of higher order. As
can be seen from the figure, such discretization effects appear to be
larger at $x_{\rm ref}=5.0$, which certainly justifies the exclusion of
those data points obtained on the two coarser lattices.

The deviation of $\zM^{\rm{w}}(r_0\mw)$ from its continuum value
amounts to about 20\% at $\beta=6.0$ and 40\% at $\beta=5.85$. Thus,
if $\zM$ were evaluated using \eq{eq_zMexpr} with $g_0^\prime=g_0$
instead of \eq{eq_zMfinal}, the result at $\beta=5.85$ would be 40\%
smaller, due entirely to cutoff effects of order~$a^2$ in the
intermediate Wilson regularization.

This discussion underlines that it is important to use the continuum
result $U_{\rm M}$ in the renormalization condition. It guarantees
that the approach to the continuum limit of quantities for which these
renormalization constants are used is not obscured by lattice
artefacts of O($a^2$), introduced by using O($a$)-improved Wilson
fermions as an intermediate regularization.

As can be inferred from \eq{eq_UM_ref}, the factor $U_{\rm M}$ is
known with a precision of about 3\%. At present this presents a lower
bound on the accuracy in the determination of renormalization factors
according to our proposal. One might argue that more precise estimates
for $U_{\rm{M}}$ could be obtained by including the data with
$\beta\leq6.0$ in the continuum extrapolation. We have still decided
against this, in order to be certain about excluding effects from
higher orders in the lattice spacing. Also, given the high cost of
simulations employing discretizations such as overlap fermions, it
will be some time before the lower bound of 3\% will be regarded as a
real limitation. If one wants to improve the accuracy in the
determination of $U_{\rm M}$, it would be preferable to add data
points at smaller lattice spacings, which should be possible with a
relatively modest amount of computer time.

\section{Lattice calculation of pseudoscalar masses \label{sec_masses}}

We now describe the details of our calculation of pseudoscalar
two-point functions using the overlap operator as well as
O($a$)-improved Wilson fermions at $\beta=5.85$. Although our strategy
only requires published results for Wilson fermions, the additional
Wilson calculations performed for different lattice sizes provide us
with valuable information about finite-size effects at a much lower
cost than with overlap fermions.

\subsection{Simulation details} 

For the massive overlap operator $D_N$ we use the following definition
\be
   D_N = \left(1-\frac{a\mov}{2(1+s)}\right)D_N^{(0)} + \mov,
\label{dn}
\ee
where
\begin{equation}
   aD_N^{(0)} =  (1+s) \left( 1 - A/ \sqrt{A^\dagger A}\right),
   \quad A\equiv 1+s-aD_W,
\end{equation}
and $D_W$ is the standard Wilson-Dirac operator. The naive continuum
limit of $D_N$ is the canonically normalized Dirac operator with a
bare quark mass $\mov$. The parameter~$s$ was fixed at $s=0.6$, which
is close to the value where the localization of the operator was found
to be optimal at this $\beta$ value \cite{hjl}. The numerical
implementation of the inverse square root in eq.~(\ref{dn}) was
performed using a Chebyshev approximation, and for further details we
refer to our earlier work~\cite{cond:paperI}. Here we only mention
that we have employed a multi-mass solver~\cite{multimass}, which
allows for a simultaneous calculation of the quark propagators for
seven values of the bare mass ranging from $a\mov = 0.047 - 0.188$.

In order to compute pseudoscalar meson masses and current quark masses
for O($a$)-improved Wilson fermions we have followed the same
procedure as in ref.~\cite{mbar:pap2}. In particular, we have employed
Schr\"odinger functional boundary conditions on lattice sizes $L/a=8,
10, 12$ and~16, with $T/a=24$ fixed, in order to study finite-size
effects.\footnote{As an orientation for the reader we add that
$L/a=12$ corresponds to $L\approx1.5\,\fm$.} For the improvement
coefficients $\csw$ and $\ca$, which appear in the definition of the
improved action and axial current, respectively, we have chosen
\be
   \csw=1.909,\qquad\ca=-0.144.
\label{eq_csw_ca}
\ee
The value for $\csw$ was obtained from the interpolating formula
eq.~(4) of ref.~\cite{impr:SCRI}. To our knowledge a non-perturbative
result for the coefficient $\ca$ has so far not been published for
$\beta<6.0$. The choice in eq.~(\ref{eq_csw_ca}) is based on a
non-perturbative calculation using the SF, which is described in
detail in Appendix~\ref{app1}.

While it can be shown that there are no exceptional configurations for
the overlap operator at non-zero quark
mass~\cite{Hasenfratz:1998ri,chiral:ovlp,Martin_exact}, this is not
the case for Wilson fermions. Indeed, when working below $\beta=6.0$
the incidence of exceptional configurations may be so high --
especially for quark masses below that of the strange quark -- so as
to make a reliable calculation of quark and meson masses impossible.

For our calculations of quark propagators using Wilson fermions we
have chosen relatively heavy quarks, with unrenormalized current quark
masses in the range $80-120\,\MeV$. Only on the smaller volumes of
$L/a=8,\,10$ did we push to lighter quarks, corresponding to
65\,MeV. We checked against the occurrence of exceptional
configurations by plotting the Monte Carlo history of the correlation
functions of the axial current and pseudoscalar density evaluated at
$x_0=T/2$. Exceptional configurations manifest themselves as isolated
peaks (or dips) whose heights exceed the typical statistical
fluctuations by several orders of magnitude. Although one
candidate each was detected for $L/a=8$ and~10, the observed
fluctuations were not deemed large enough to justify the exclusion of
these configurations from the statistical ensembles. Hence, for the
case of Wilson fermions, all hadron and quark masses have been
evaluated using the full statistics on all lattices.

\subsection{Results for pseudoscalar masses}
\label{sec:respsmasses}

\begin{table}[t]
\centering
\begin{tabular}{r c c c c c}
\hline
\hline \\[-1.0ex]
  $L/a$ & configs. & $\kappa$ & $a\mw$      & $a\mps$    & $(a\mps)^2$
\\[1.0ex]
\hline \\[-1.0ex]
    8   &   640    & 0.13150  & 0.07309(37) & 0.5780(45) & 0.3342(52) \\
        &          & 0.13200  & 0.06106(40) & 0.5265(49) & 0.2772(51) \\
        &          & 0.13250  & 0.04893(47) & 0.4707(54) & 0.2216(50) \\
        &          & 0.13300  & 0.03685(52) & 0.4085(60) & 0.1668(49) \\
\hline
   10   &   512    & 0.13150  & 0.07362(27) & 0.6002(29) & 0.3602(35) \\
        &          & 0.13200  & 0.06174(28) & 0.5496(31) & 0.3021(35) \\
        &          & 0.13250  & 0.04991(30) & 0.4948(35) & 0.2449(35) \\
        &          & 0.13300  & 0.03805(34) & 0.4338(40) & 0.1882(35) \\
\hline
   12   &   256    & 0.13150  & 0.07436(30) & 0.6005(30) & 0.3606(36) \\
        &          & 0.13200  & 0.06249(32) & 0.5503(33) & 0.3028(36) \\
        &          & 0.13250  & 0.05067(35) & 0.4961(36) & 0.2461(36) \\
\hline
   16   &   150    & 0.13150  & 0.07458(24) & 0.6065(21) & 0.3678(26) \\
        &          & 0.13200  & 0.06277(25) & 0.5573(23) & 0.3106(26) \\
        &          & 0.13250  & 0.05102(27) & 0.5045(26) & 0.2545(26) \\[1.0ex]
\hline
\hline
\end{tabular}
\caption{\it Lattice sizes, statistics, pseudoscalar and current
quark masses computed using O(a)-improved Wilson fermions.}
\label{tab:pion_w}
\end{table} 
\begin{table}[t]
\centering
\begin{tabular}{ccc}
\hline
\hline
  $a\mov$ & $a\mps$   & $(a\mps)^2$ \\
\hline
  0.047   & 0.311(31) &  0.096(20) \\
  0.063   & 0.357(22) &  0.128(16) \\
  0.078   & 0.398(17) &  0.159(14) \\
  0.097   & 0.444(14) &  0.197(13) \\
  0.125   & 0.506(12) &  0.256(13) \\
  0.161   & 0.579(11) &  0.335(12) \\
  0.188   & 0.630(10) &  0.397(12) \\
\hline
\hline
\end{tabular}
\caption{\it 
Pseudoscalar masses extracted from single-mass fits in the interval
$6\leq x_0/a\leq12$ at several values of the bare mass
$\protect\mov$.}
\label{tab:pion_ov}
\end{table} 

The results for the pseudoscalar and bare current quark masses
computed for O($a$)-improved Wilson fermions are presented in
Table~\ref{tab:pion_w}. The masses were extracted following the
procedure described in detail in ref.~\cite{mbar:pap2}. In particular,
the estimates for the pseudoscalar masses listed in the table were
obtained by averaging the effective masses computed from the
correlation function of the improved axial current. In accordance with
Table~1 of~\cite{mbar:pap2}, and using
$r_0/a=4.067\pm0.014$~\cite{pot:r0_SU3} we have chosen the time window
$11\leq x_0/a\leq15$ for the averaging procedure. Estimates for the
bare current quark mass were obtained in a similar manner, by
averaging the results over the interval $10\leq x_0/a\leq16$. This
choice of time window coincides with a clear plateau observed for this
quantity.

By comparing the results in Table~\ref{tab:pion_w} obtained on
different lattice sizes, one observes that the time interval $11\leq
x_0/a\leq15$ suggested in~\cite{mbar:pap2} produces consistent results
for pseudoscalar masses if $L/a\geq10$. By contrast, using a spatial
lattice size of $L/a=8$ (corresponding to $L\approx1\,\fm$) leads to
finite-size effects at the level of 4\% (4 standard
deviations). Choosing a shorter time window on $L/a=8$, such as
$11\leq x_0/a\leq13$, brings the results closer to those on the larger
volumes, but a 2\% effect (1.5 standard deviations) remains.

On the other hand, for $L/a\geq10$ (i.e. $L\geq1.2\,\fm$) finite-size
effects in the pseudoscalar mass are of the order of 1.5\% or less. In
order to exclude large finite-size effects for pseudoscalar masses
computed using the overlap operator, whilst keeping the computational
overheads small, we have thus decided to work with $L/a=10$. However,
the above analysis of finite-size effects only applies to pseudoscalar
meson masses with $a\mps\;\gtaeq\;0.5$, a range which lies above the
reference points $x_{\rm ref}=1.5736$ and~3.0. This was our main
reason to consider the reference point $x_{\rm ref}=5.0$ in addition:
at $\beta=5.85$ this choice corresponds to $a\mps\approx0.55$, a value
for which the absence of significant finite-volume effects has been
confirmed.

In Table~\ref{tab:pion_ov}, we present the results for pseudoscalar
masses computed using the overlap operator on an ensemble of 50
configurations. The chosen lattice size was $L/a=10,\,T/a=24$, with
periodic boundary conditions in all space-time directions. After
averaging the correlation functions over the forward and backward
directions in euclidean time, we extracted the masses from single-cosh
fits to the correlation functions in the interval $6\leq
x_0/a\leq12$. The quoted statistical errors were obtained from a
jackknife procedure. We have classified configurations
according to their topological index, distinguishing between
topological (having a non-zero index) and non-topological
configurations and determined the average pseudoscalar masses
restricted to either class. Only very near the chiral limit does one
expect these quantities to differ. In the range of quark masses
considered we did not detect any difference between the two classes,
so that we could safely include all topologies in the average.

\begin{figure}[t]
\centering
\vspace{-5.7cm}
\epsfig{file=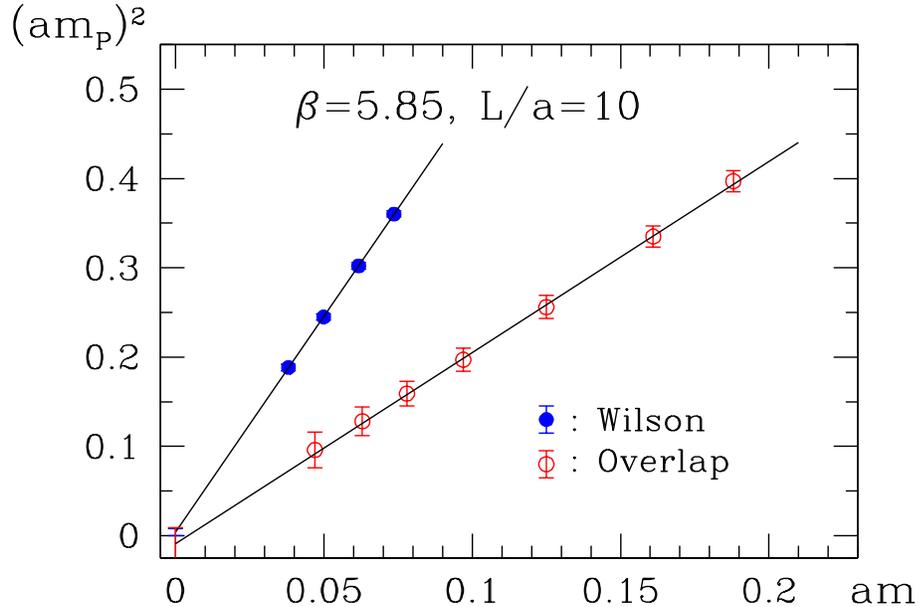, width=16cm} 
\vspace{-2.0cm}
\caption{\it $\mps^2$ versus $\mov$ (open circles) and $\mw$ (filled
circles) respectively for overlap and $O(a)$-improved Wilson
fermions. The solid lines are the fits of eqs.~(\ref{mpivsmov}) and
(\ref{mpivsmw}).}
\label{fig_mps_vs_m}
\end{figure}

In Fig.~\ref{fig_mps_vs_m}, we show the results for the behaviour of
$(a\mps)^2$ as a function of $a\mov$ and $a\mw$. In both cases the
results are perfectly compatible with the linear behaviour expected in
lowest order Chiral Perturbation Theory. As an illustration we have
plotted the results from a chiral fit to our data, using the linear
parameterization
\be
   (a\mps)^2 = A_i + B_i\cdot(am_i),\quad i=\hbox{ov,\,w},
\label{eq_mps2vsm}
\ee
thus allowing for a non-zero intercept of $(a\mps)^2$ at vanishing
quark mass. The results of these fits are:
\bea
   A_{\rm ov} = -0.009(18),  &\quad & B_{\rm ov} = 2.139(80),
   \label{mpivsmov} \\
   A_{\rm w}  = \phantom{-}
                 0.0039(39), &\quad & B_{\rm w}  = 4.836(44),
   \label{mpivsmw}
\eea
where we have used all seven data points in the overlap case and all
four quark masses in the Wilson set. Note that in both cases the
intercepts are perfectly compatible with zero within errors.

Pseudoscalar propagators for overlap fermions have been studied before
in~\cite{Pions:liu99} for $\beta=5.7, 5.85$ and~6.0, thus enabling a
direct comparison with our results.\footnote{Note that the bare mass in
\cite{Pions:liu99} differs from $\mov$. To leading order 
in $a$ the two definitions differ by a factor $(1+s)$.}. Since the
authors of \cite{Pions:liu99} do not present numbers for $(a\mps)^2$
one has to infer its quark mass behaviour by reading the data off
their plots. From this fairly crude comparison we conclude that the
slope parameter $B_{\rm ov}$ is in rough agreement with our
findings. Unlike the authors of \cite{Pions:liu99} we did not attempt
to model the quark mass behaviour including quenched chiral
logarithms. Judging from the quality of our fits to \eq{eq_mps2vsm} we
do not expect a significant deviation from leading order Chiral
Perturbation Theory at our level of statistics in the range of quark
masses we consider. In order to test for the presence of quenched
chiral logarithms, it may be necessary to go to much smaller masses
than our lightest quark, whose mass is roughly half as large as that
of the strange quark.

\section{Results and discussion \label{sec_results}}

\subsection{Non-perturbative result for $\zshat$}

We have now the necessary results to compute the renormalization
factor $\zshat$ for the RGI condensate using eqs.~(\ref{eq_zMfinal})
and~(\ref{eq_zs_zm}). Our task is to determine $(r_0\mov)$ at the
reference value $x_{\rm ref}$, which could in principle be achieved
using the results of the chiral fit, \eq{mpivsmov}. However, we prefer
to perform local interpolations of the quark mass $\mov$ to the
reference point, which avoids any assumption about the mass behaviour
of $(a\mps)^2$ at or very near the chiral limit.

To this end we have interpolated $(r_0\mov)$ to our chosen values of
$x_{\rm{ref}}$ by using the three nearest data points for
$(r_0\,\mps)^2$. We obtain
\be
   (r_0\,\mov)\Big|_{(r_0\mps)^2=x_{\rm ref}} =\left\{
   \begin{array}{cl}
      0.190(44),&\quad x_{\rm ref}=1.5736 \\
      0.363(26),&\quad x_{\rm ref}=3.0 \\
      0.594(23),&\quad x_{\rm ref}=5.0.
   \end{array} \right.
\label{eq_r0mov_res}
\ee
We add that these results are entirely consistent with interpolations
using only the two neighbouring points. Given the almost perfect
linearity of the quark mass dependence of $(r_0\mps)^2$, it is not
surprising that interpolations using all seven data points are also
consistent. Hence we regard the
uncertainties in our results of \eq{eq_r0mov_res} as conservative
estimates. 

By combining the results for $(r_0\,\mov)$ with the factor $U_{\rm M}$
we obtain
\be
   \zshat =\left\{
   \begin{array}{ll}
      1.05(25),&\quad x_{\rm ref}=1.5736 \\
      1.04(8), &\quad x_{\rm ref}=3.0 \\
      1.02(4), &\quad x_{\rm ref}=5.0,
   \end{array} \right. \quad \beta=5.85.
\label{eq_zshat_res}
\ee
The results at the three values of $x_{\rm ref}$ are consistent within
even the smallest of the statistical errors. This indicates that our
renormalization condition may be applied over a fairly large range of
quark masses without introducing significant discretization errors due
to working at non-zero quark mass. It also indicates that
finite-volume effects, even at the lightest reference point, appear to
be small. Owing to the modest statistics in our simulations, the
errors in $\zshat$ in~\eq{eq_zshat_res} are clearly dominated by the
limited accuracy of the estimates for $(r_0\,\mov)$ at the reference
points, which amounts to 24\%, 8\% and 4\% at $x_{\rm ref}=1.5736$,
3.0 and~5.0, respectively. This is larger than the relative error in
$U_{\rm M}$ of about 3\%. However, with better statistics, it should
be easy to obtain more accurate estimates for $(r_0\,\mov)$ and thus
$\zshat$.

The dependence on the value of $x_{\rm ref}$ is quite weak and well
covered by the statistical uncertainty. We quote the result for
$x_{\rm ref}=3.0$ as our best estimate, since this reference point is
a good compromise between being chiral enough so as not to introduce
significant discretization errors, and massive enough to guarantee
negligible finite-size effects. Thus we obtain
\be
   \begin{array}{rl}
   \zshat       & = 1.04\pm0.08 \\
   \zs(2\,\GeV) & = 1.44\pm0.11,
   \end{array}\quad \beta=5.85,
\label{eq_zs_res}
\ee
These numbers are the main result of this paper. All systematics
associated with the intermediate Wilson regularization have been
eliminated by the continuum extrapolation in the definition of the
universal factor $U_{\rm M}$. As discussed in Section
\ref{sec_masses}, we have verified that finite-volume effects are
under control. Discretization errors of order $(am)^2$ in our
determination of $\zshat$ seem to be small, as indicated by the
consistency of our results at the three reference values.

\subsection{Comparison with perturbation theory}

In ref.~\cite{chiral:AlFoPaVi} the perturbative renormalization of
quark bilinears was studied for the overlap operator. The one-loop
expression for $\zs$ reads
\be
   \zs^{\rm pt}(g_0,\mu) = 1+g_0^2\left[\frac{1}{2\pi^2}\ln(a\mu) 
                 +z_{\rm S}^{(1)}\right]+O(g_0^4),
\label{eq_zspbare}
\ee
where the one-loop coefficient $z_{\rm S}^{(1)}$ depends on the
parameter~$s$ in the definition of the overlap operator. For our value
of~$s=0.6$ one finds~\cite{chiral:AlFoPaVi}
\be
   z_{\rm S}^{(1)}=0.107074.
\ee
It is well known that perturbation theory in the bare coupling~$g_0$
is not very convergent. As a consequence it has become customary to
consider ``mean-field improved'' estimates for perturbative
renormalization factors~\cite{Parisi,lepenzie93}. Another proposal is
based on the resummation of ``cactus''
diagrams~\cite{chiral:AlFoPaVi}. Here, in the spirit of
\cite{lepenzie93}, we propose the following mean-field improved
expression:
\be
   \zs^{\rm mf}(g_0,\mu) = \left(\frac{1+s}{1+\tilde s}\right)
\left\{ 1+g^2
      \left[\frac{1}{2\pi^2}\ln(a\mu)+z_{\rm S}^{(1)}+u_0^{(1)}\left(
\frac{3-s}{1+s}\right)
      \right]\right\}\ ,
\label{eq_zspttad}
\ee
where $\tilde s=3+(s-3)/u_0$. We take $u_0^4$ to be the average
plaquette in infinite volume, and $u_0^{(1)}=-1/12$ is the one-loop
coefficient in its perturbative expansion. At this order, the choice
of coupling, $g$, is ambiguous. For consistency, we work with the
coupling used in obtaining the renormalization group factor
in~\eq{eq_mbaroverM}, i.e. $g^2=2.5432$. We can now evaluate
eq.~(\ref{eq_zspttad}) using lattice data for the average plaquette
and by setting the lattice spacing at $\beta=5.85$ with the help of
the interpolating formula for the hadronic
radius~$r_0/a$~\cite{pot:r0_SU3}. For $u_0^4=0.575$ the results for
$\zshat^{\rm mf}$ and $\zs^{\rm mf}$ are
\be
   \begin{array}{rl}
   \zs^{\rm mf}(2\,\GeV) & = 1.26 \\
   \zshat^{\rm mf}       & = 0.91,
   \end{array}\quad \beta=5.85.
\label{eq_zs_mf_res}
\ee
Comparing these mean-field improved perturbative estimates to the
non-perturbative results of \eq{eq_zs_res}, one finds that the former
are about 12\% smaller. This is a significant improvement on the
results of bare perturbation theory, where $\zs^{\rm
pt}(2\,\GeV)=1.12$ as given by \eq{eq_zspbare}, which are more than
20\% smaller than the non-perturbative results of \eq{eq_zs_res}.

\section{Summary and conclusions \label{sec_concl}}

We have proposed and tested a method to non-perturbatively renormalize
scalar and pseudoscalar densities for fermionic discretizations which
preserve chiral symmetry at non-zero lattice spacing. This method also
provides a means to renormalize quark masses non-perturbatively in any
lattice regularization. Given the cost of simulations employing
Ginsparg-Wilson fermions and the incompatibility of Schr\"odinger
functional boundary conditions with the Ginsparg-Wilson equation, we
found it advantageous to proceed through O($a$)-improved Wilson
fermions, where the non-perturbative renormalization of currents and
densities has already been studied extensively. The sought-after
renormalization constants are then obtained through a matching of RGI
quark masses at fixed pseudoscalar meson mass. As evident in
\eq{eq_zMfinal}, all reference to Wilson fermions drops out, owing to the
universal factor $U_{\rm M}$ defined and evaluated in the continuum
limit. Furthermore, the overhead for implementing the matching
condition should be negligible in most cases, since it relies on
quantities that are commonly computed in phenomenological studies
where the renormalization constants are likely to be used. The idea to
use an intermediate lattice regularization which is relatively cheap
to implement may prove useful for the computation of other
renormalization constants in fermionic discretizations that are
numerically much more demanding.

As an application, we have computed the renormalization constants
which are required for the quark condensate obtained using overlap
fermions in the quenched approximation. Our results at $\beta=5.85$
are listed in \eq{eq_zs_res} and can now be combined with the result
for the subtracted bare condensate determined in \cite{cond:paperI},
i.e. $a^3\Ssub=0.00323(37)$~\footnote{In ref.~\cite{cond:paperI} the
massive overlap operator was not O($a$) improved. The above result for
$a^3\Ssub$ was obtained for the O($a$) improved definition in
\protect\eq{dn}, which amounts to a redefinition of the tree level
mass. The difference from the result quoted in \cite{cond:paperI} is
less than 1\%.}. In units of $r_0$ we obtain
\be
   \begin{array}{rl}
   r_0^3\SRGI                  & = 0.226(26)(2)(17) \\
   r_0^3\Sigma_\msbar(2\,\GeV) & = 0.313(36)(3)(23),
   \end{array}\quad \beta=5.85,
\label{eq_cond_res_r0}
\ee
where the first error is due to the statistical uncertainty in
$a^3\Ssub$, the second corresponds to the error in $r_0/a$, and the
third arises from the error in the renormalization factors $\zshat$
and $\zs$, respectively. Combining all but the error associated with
$r_0/a$ in quadrature and using $r_0=0.5\,\fm$ we find
\be
   \begin{array}{rl}
   \SRGI                  & = \left(240\pm11\,\MeV\right)^3 
   \times
   \left(\displaystyle\frac{a^{-1}[\MeV]}{1605\,\MeV}\right)^3\\
   \Sigma_\msbar(2\,\GeV) & = \left(268\pm12\,\MeV\right)^3
   \times
   \left(\displaystyle\frac{a^{-1}[\MeV]}{1605\,\MeV}\right)^3
   \end{array}\quad \beta=5.85.
\label{eq_cond_res_mev}
\ee
These results are still subject to discretization errors of O($a^2$).
A detailed discussion of lattice artefacts and other systematic errors
-- including the scale ambiguity in the quenched approximation -- as
well as a comparison of these results with those obtained through
other approaches is deferred to our companion
paper~\cite{cond:paperIV}.

\vspace{1cm}
\par\noindent
{\bf Acknowledgements.}
\par\noindent
We are grateful to Martin L\"uscher, Stefan Sint and Rainer Sommer for
useful and stimulating discussions. We thank the computer centres at
NIC (J\"ulich) and DESY Zeuthen for providing computer time and
technical support.


\begin{appendix}
\section{Determination of $\ca$ at $\beta=5.85$ \label{app1}}

\begin{figure}[t]
\centering
\vspace{-7.3cm}
\epsfig{file=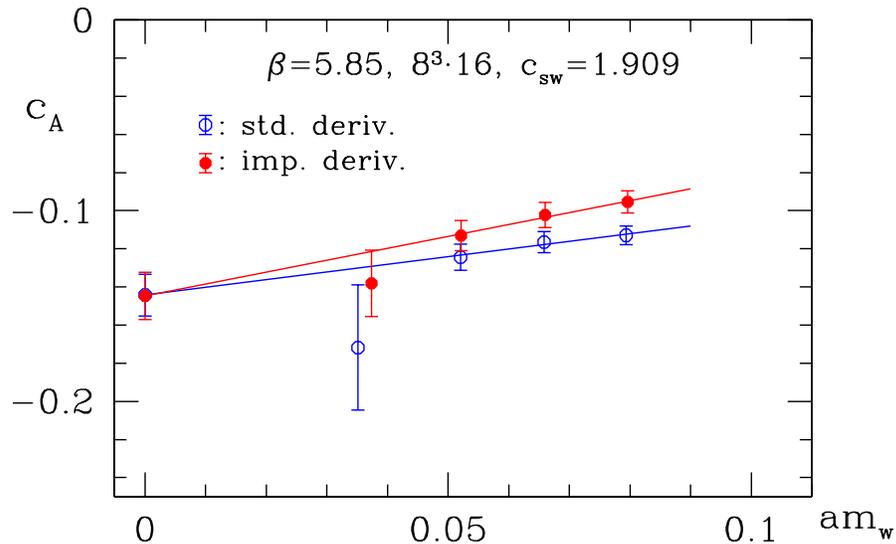, width=16cm}
\vspace{-2.0cm}
\caption{\it The chiral extrapolation of $\ca$ using the three
heaviest quark masses. Data for the standard and improved definitions
of the lattice derivative are denoted by the open and filled symbols,
respectively.\label{fig_ca_ext}}
\end{figure}

\begin{figure}[t]
\centering
\vspace{-8.5cm}
\epsfig{file=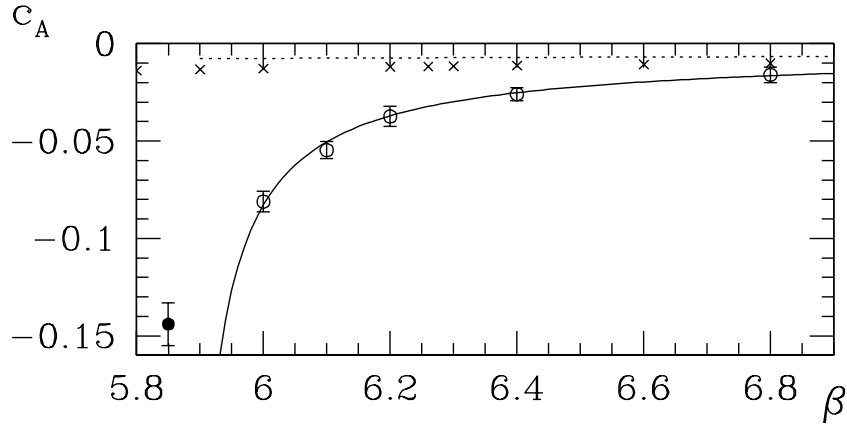, width=16cm}
\vspace{-2.0cm}
\caption{\it The result for $\ca$ at $\beta=5.85$ (filled circle)
compared with the results from ref.~\protect\cite{impr:pap3} (open
circles). The solid curve is the interpolating formula valid for
$6.0\leq\beta$. The dashed curve denotes perturbation theory in the
bare coupling, whereas the crosses represent the estimates in
mean-field improved perturbation theory. \label{fig_ca_all}}
\end{figure}

While a non-perturbative value for the improvement coefficient
$\csw$ is available at $\beta=5.85$~\cite{impr:SCRI}, the coefficient
$\ca$ of the improved axial current, which is required for the current
quark mass, is not known for $\beta<6.0$.

As mentioned in \sect{sec_masses}, numerical simulations using Wilson
fermions are hampered by the occurrence of exceptional configurations
associated with unphysical zero modes of the Wilson-Dirac
operator. This problem is further exacerbated by working at
$a\;\gtaeq\;0.1\,\fm$ (i.e. $\beta\;\lesssim\;6.0 $) and small quark
masses. However, if one is willing to relax the requirement that the
improvement conditions for $\csw$ and $\ca$ be evaluated at or very
near the chiral limit~\cite{impr:pap3}, the occurrence of exceptional
configurations may be sufficiently suppressed. Thus, an extension of
the determination of improvement coefficients to the regime where
$\beta<6.0$ may be possible.

In order to determine $\ca$ at $\beta=5.85$ we have chosen
$\csw=1.909$~\cite{impr:SCRI} and followed the strategy of
ref.~\cite{impr:pap3}. It has been observed~\cite{RS_MG_priv}, though,
that the original improvement condition for $\ca$ used
in~\cite{impr:pap3} suffers from a loss of numerical accuracy towards
larger values of the bare coupling, corresponding to
$\beta\approx6.0$. A modified, but closely related improvement
condition for $\ca$ has been proposed and tested~\cite{RS_MG_priv}. It
was shown that the alternative condition's numerical sensitivity does
not deteriorate at large couplings. Furthermore, it gives consistent
results compared with the original condition, thereby confirming the
determination of $\ca$ in ref.~\cite{impr:pap3}.

We have evaluated both the original and modified improvement condition
at $\beta=5.85$ on a $8^3\cdot16$ lattice. We used the same four
values of the hopping parameter listed in Table~\ref{tab:pion_w}. With
the procedure outlined in~\sect{sec_masses} to check against the
occurrence of exceptional configurations, we eliminated two
configurations from the original ensemble of~2560. It was found that
the alternative improvement condition gave a stable signal for $\ca$
at the three heaviest quark masses. By contrast, the original
improvement condition used in~\cite{impr:pap3} performed so badly at
$\beta=5.85$ that it could not be used to determine $\ca$.

Our estimates for $\ca$ obtained at non-zero quark masses shows some
dependence on the quark mass, and therefore our final result is
obtained through a chiral extrapolation as shown in
Fig.~\ref{fig_ca_ext}. We have also evaluated the improvement
condition using a higher-order lattice derivative~\cite{impr:roma2_1},
which was found to have a significant impact on the determination of
some improvement coefficients~\cite{impr:bAbP}. The corresponding
results and extrapolation are also shown in
Fig.~\ref{fig_ca_ext}. Both definitions of the lattice derivative give
entirely consistent results in the chiral limit, and as our final
result we quote
\be
   \ca=-0.144\pm0.011,\qquad\beta=5.85.
\label{eq_ca_result}
\ee
This value is shown together with the previous determination of $\ca$
for $\beta\ge6.0$~\cite{impr:pap3} in Fig.~\ref{fig_ca_all}.

Finally we note that we have also estimated $\ca$ using the method
proposed in ref.~\cite{impr:lanl1} (see
also~\cite{impr:lanl2,impr:sara_lat00}). We have confirmed the
observation of~\cite{impr:sara_lat00}, namely that the accuracy of the
method is limited by the relatively small range in $x_0$ in which a
stable signal is obtained. On our $8^3\cdot16$ lattice
the method of~\cite{impr:lanl1} produces results which are entirely
consistent with the estimate in eq.~(\ref{eq_ca_result}).

\clearpage

\end{appendix}
\bibliography{biblist,newrefs}        
\bibliographystyle{h-elsevier}   
\end{document}